\documentclass[lettersize,journal]{IEEEtran}
\usepackage{amsmath,amsfonts}
\usepackage{algorithm}
\usepackage{array}
\usepackage[caption=false,font=normalsize,labelfont=sf,textfont=sf]{subfig}
\usepackage{textcomp}
\usepackage{stfloats}
\usepackage{url}
\usepackage{verbatim}
\usepackage{graphicx}
\usepackage{cite}
\usepackage{tabularx}
\usepackage{quantikz}
\usepackage{booktabs}
\usepackage{amssymb,amsthm}
\usepackage[hidelinks]{hyperref}
\usepackage{bm}
\usepackage{algpseudocode}
\usepackage{makecell}
\usetikzlibrary{arrows.meta}
\newtheorem{theorem}{Theorem}[]

\usepackage{array}
\usepackage{placeins}
\newcolumntype{L}[1]{>{\raggedright\arraybackslash}p{#1}}
\begin{document}

\title{A Quantum Optimization Algorithm for Optimal Electric Vehicle Charging Station Placement for Intercity Trips}

\author{
    Tina Radvand, Alireza Talebpour, and Homa Khosravian%
    \thanks{Tina Radvand and Alireza Talebpour (corresponding author) are with The Grainger College of Engineering, University of Illinois at Urbana-Champaign, Urbana, IL 61801 USA (e-mail: radvand2@illinois.edu; ataleb@illinois.edu).}
    \thanks{Homa Khosravian is with the National Science Foundation, Alexandria, VA 22314 USA (e-mail: hkhosrav@nsf.gov).}
    \thanks{Funding for this research was provided by the Center for Connected and Automated Transportation under Grant No. 69A3552348301 of the U.S. Department of Transportation, Office of the Assistant Secretary for Research and Technology (OST-R), University Transportation Centers Program.}
}

\markboth{Journal of \LaTeX\ Class Files,~Vol.~14, No.~8, August~2021}%
{Shell \MakeLowercase{\textit{et al.}}: A Sample Article Using IEEEtran.cls for IEEE Journals}


\maketitle

\begin{abstract}
Electric vehicles (EVs) play a significant role in enhancing the sustainability of transportation systems. However, their widespread adoption is hindered by inadequate public charging infrastructure for long-distance travel. Identifying optimal charging station locations in large transportation networks is an NP-hard combinatorial optimization problem. This paper applies Grover Adaptive Search (GAS) to improve the efficiency of solving the Charging Station Location Problem (CSLP). The proposed method achieves a quadratic improvement in computational complexity over classical exact methods, such as branch and bound. This paper develops a quantum subroutine that encodes the CSLP constraints by marking feasible solutions with objective value below a given threshold, and integrates this subroutine within the GAS procedure. The approach is demonstrated on a 7-node transportation network in central Illinois, with an analysis of success probability and sensitivity to algorithm parameters.
\end{abstract}

\begin{IEEEkeywords}
Electric Vehicle, Charging Station Location, Grover Adaptive Search, Quantum Optimization.
\end{IEEEkeywords}

\section{Introduction}

According to the 2023 inventory of U.S. greenhouse gas emissions and sinks~\cite{epa2023inventory}, the transportation sector produced the largest share of U.S. greenhouse gas emissions in 2021. As a result, emission regulations have become stricter~\cite{change2007intergovernmental}, encouraging a shift toward more sustainable transportation alternatives.

Electric vehicles (EVs) can effectively address sustainability concerns in the transportation sector~\cite{del2018life}. They produce zero tailpipe emissions~\cite{nealer2015review}, have high energy conversion efficiency~\cite{wen2020overview}, and operate with lower noise levels than internal combustion engines (ICEs)~\cite{campello2017effect}. EVs also improve energy security by reducing dependence on imported energy sources~\cite{fontaine2008shortening}. Moreover, within the smart charging framework, EVs can potentially play a dual role by supporting the stability of the renewable energy grid~\cite{nunes2015enabling} and providing power during grid stress through vehicle-to-grid capabilities~\cite{khan2018analyzing}.

While EVs offer numerous advantages, several concerns persist, limiting their broader adoption. As of 2022, EVs accounted for less than $1\%$ of vehicles in the United States~\cite{AFDCVehicleRegistration}. The slow market penetration of EVs can be attributed to several factors, including their high purchase cost~\cite{carley2013intent}, long charging time~\cite{hackbarth2013consumer}, short driving range (compared with ICE vehicles)~\cite{dimitropoulos2013consumer}, and the limited or poorly distributed public charging infrastructure~\cite{gass2014analysis}. The driving range offered by a fully charged battery of an EV is generally sufficient for the everyday needs of the majority of users. For example, in the U.S. approximately $95\%$ of trips do not exceed $120$ miles~\cite{krumm2012people}, whereas the average EV range is around $270$ miles~\cite{fueleconomygov}. However, longer intercity trips cannot be completed on a single charge and necessitate the availability of public charging infrastructure on interstate highways. Supporting such trips is also critical for long-haul freight, as electric trucks are expected to become more widespread.  

For long-distance trips, drivers often perceive charging time as ``deadtime'' \cite{graham2012mainstream}, highlighting the importance of installing fast chargers to support such travel. However, installing and relocating fast charging stations is costly \cite{jochem2016optimizing}. Therefore, the optimal locations of charging infrastructure should be carefully identified to ensure widespread access to EV owners.

Despite extensive research on the charging station location problem (CSLP)~\cite{kchaou2021charging}, it remains NP-hard~\cite{lam2014electric}. Exact methods, such as branch and bound, become computationally impractical for large or complex networks due to the exponential growth of the solution space~\cite{bao2021optimal}. As a result, many studies rely on heuristic and metaheuristic approaches. While these methods can efficiently produce feasible solutions, they do not guarantee optimality and may miss the global optimum due to their approximate and iterative nature.

This study applies Grover Adaptive Search (GAS)~\cite{durr1996quantum} to the CSLP by developing two problem-specific quantum subroutines. The first subroutine identifies combinations of charging station locations that allow all trips to be completed without vehicles depleting their battery charge. The second subroutine evaluates the cost of candidate combinations and marks those with objective value below a specified threshold. Together, these subroutines formulate the CSLP within the GAS framework. The computational complexity of the proposed formulation is $\mathcal{O}(1.4^{|\mathcal{N}|})$, compared with $\mathcal{O}(2^{|\mathcal{N}|})$ for classical exact methods, where $|\mathcal{N}|$ denotes the number of candidate charging station locations.

Existing quantum approaches to charging infrastructure planning have primarily relied on heuristic paradigms, which do not provide optimality guarantees. In contrast, the use of GAS enables an exact optimization framework with provable complexity bounds. To the best of our knowledge, this work is the first application of a Grover-based quantum optimization framework to the CSLP.

The remainder of this paper is organized as follows. The \textbf{Literature Review} section surveys classical approaches to the CSLP and reviews quantum optimization techniques, including prior quantum approaches applied to the CSLP. The \textbf{Problem Statement} section formally defines the CSLP, along with the modeling assumptions and research objectives. The \textbf{Complexity Analysis} section presents a proof of NP-hardness for the CSLP. The \textbf{Background on Grover-Based Optimization} section reviews the Grover search algorithm and its extension, Grover Adaptive Search. The \textbf{Quantum Circuit Design} section describes the problem-specific quantum subroutines and circuit constructions used to formulate the CSLP within the GAS framework. The \textbf{Analysis and Results} section applies the proposed method to a transportation network in central Illinois and evaluates its performance. Finally, the \textbf{Conclusion and Future Work} section summarizes the findings and outlines directions for future research.

\section{Literature Review}

The growing interest in accelerating EV adoption has motivated a substantial body of literature on the CSLP. Unlike facility location problems, where the demand is expressed as weights at network nodes, most CSLP formulations represent demand as flows between origin--destination pairs. Charging stations must be placed so that the distance between consecutive stations is within the driving range of EVs.

Hodgson \cite{hodgson1990flow} introduced the concept of flow-capturing demand in location--allocation models. His work focused on placing facilities to maximize demand coverage, where a demand is considered covered if at least one facility lies along the path between an origin and a destination. However, the presence of a single charging facility along a route is insufficient to ensure that an electric vehicle can complete the trip without depleting its battery. 

Recognizing this limitation, Kuby and Lim \cite{kuby2005flow} proposed the flow-refueling location model (FRLM), which models the CSLP by generating feasible combinations of charging station locations in a preprocessing step to ensure trip feasibility. In the FRLM, demand is considered captured if at least one feasible combination supports travel between the origin and destination, and the objective is to maximize the total captured demand. A key limitation of the FRLM is the computational cost of generating feasible station combinations, which often necessitates the use of heuristic methods \cite{capar2012efficient}. As a result, several alternative approaches have been proposed to solve the CSLP.

One such approach is the arc-covering model introduced by Capar \cite{capar2012efficient}, in which a trip is considered covered only if every arc along the route is covered by a charging station. Another approach was proposed by MirHassani and Ebrazi \cite{mirhassani2013flexible}, who expanded the network by constructing path segments between origins and destinations and placing charging stations at the segment endpoints. Trip feasibility is ensured by requiring the length of each segment to remain within the driving range of the EV.

Ghamami et al. \cite{ghamami2016general, soltanpour2026supporting} proposed a general corridor model formulated as a mixed-integer program with nonlinear constraints and solved it using a simulated annealing--based metaheuristic. Their objective was to minimize total system cost while accounting for queuing delays. Bao et al. \cite{bao2021optimal} studied a bi-level CSLP under a limited budget, in which vehicles select self-optimal routes while the planner minimizes overall network congestion caused by detours. They applied both branch-and-bound and nested partitions algorithms, and observed that branch-and-bound performs well only when driving ranges are short.

The CSLP is NP-complete under all modeling approaches \cite{lam2014electric} and is commonly solved using heuristic and metaheuristic methods. Kchaou-Boujelben \cite{kchaou2021charging} has provided a comprehensive review and comparison of existing solution methods. In practice, reducing computational time often requires trading off solution quality. This has led to a shift in computing paradigms toward quantum optimization methods for the CSLP.

Quantum optimization is a rapidly growing area of quantum computing research, and several quantum optimization methods exist for combinatorial problems. Quantum Annealing (QA)~\cite{kadowaki1998quantum} is applied to problems such as the traveling salesman, job-shop scheduling, transportation network design, multicommodity network flow optimization, and scenario-based stochastic time-dependent shortest path routing \cite{martovnak2004quantum,venturelli2016job,dixit2023quantum,10858600,dixit2024quantum}. Variational quantum algorithms, including the Quantum Approximate Optimization Algorithm (QAOA)~\cite{farhi2014quantum} and the Variational Quantum Eigensolver (VQE)~\cite{peruzzo2014variational}, are studied for combinatorial problems such as Max-Cut and graph coloring \cite{do2020planning}. Other quantum optimization techniques include Grover Adaptive Search (GAS)~\cite{durr1996quantum,baritompa2005grover}, which targets optimization over large unstructured search spaces \cite{gilliam2021grover}, and quantum walk--based methods, which are used for network analysis and spatial search \cite{montanaro2020quantum,montanaro2015quantum,faccin2013degree,childs2004spatial}.

Beyond optimization-focused studies, recent work has explored the use of quantum computing in the transportation domain \cite{cooper2021exploring}. Examples include the use of the quantum Fourier transform for estimating vehicle drive cycles \cite{dixit2022quantum} and survey studies highlighting the potential of quantum computing in intelligent transportation systems \cite{zhuang2024quantum}. However, only a limited number of studies have applied quantum optimization methods to the CSLP.

{\color{black}
Rao et al. \cite{rao2023hybrid} investigated the CSLP with the objective of reducing travel distance to demand locations, while accounting for both transportation demand and the power distribution system. To solve the resulting optimization problem, they employed QA and the VQE. Their numerical results showed that the VQE did not achieve notable computational speed-up, whereas QA demonstrated runtime improvements of up to approximately six times compared to a classical binary lightning search algorithm.

Chandra et al. \cite{chandra2022towards} proposed a hybrid algorithm that combines QA with genetic algorithms to solve the CSLP, with the objective of minimizing travel distance from points of interest. Their numerical experiments showed that seeding the genetic algorithm with QA solutions led to better placement quality than using either QA or genetic algorithms alone, and also resulted in faster and more stable convergence compared to a randomly seeded genetic algorithm.

Hu et al. \cite{hu2024integrating} proposed an improved quantum genetic algorithm (IQGA) for charging station placement that minimizes total social cost, defined as the combined cost of station construction and operation, user travel and waiting time, and carbon emissions. Their numerical experiments showed that the proposed method achieved lower total cost solutions compared to a standard quantum genetic algorithm and several classical metaheuristics. However, the improved solution quality came at the expense of increased computational time due to the additional simulated annealing component.

The methods used in the literature to address charging station placement, including QA, VQE, and IQGA, each have their own strengths and limitations. In the case of QA, its main strength lies in its ability to explore complex energy landscapes through quantum tunneling. Unlike simulated annealing, which requires the system to climb energy barriers, this mechanism allows the system to pass through barriers and is effective for rugged energy landscapes in combinatorial optimization problems. However, QA is a metaheuristic and does not guarantee convergence to the global optimum in practice. It does not provide polynomial-time solutions for NP-hard problems, and observed speedups are problem dependent rather than asymptotic~\cite{albash2018adiabatic}. In addition, QA shows limited robustness across runs and requires reformulating the problem as a Quadratic Unconstrained Binary Optimization (QUBO), where constraints are incorporated into the objective function through penalty terms. Poorly chosen penalties can lead to infeasible solutions or distort the energy landscape. Moreover, penalty tuning and QUBO construction are computationally expensive and can eliminate the runtime advantage of QA~\cite{quinton2025quantum}.

VQE is a hybrid quantum--classical algorithm that uses shallow quantum circuits and is compatible with noisy quantum devices. VQE can sample correlated solutions efficiently, but it is heuristic and its convergence depends on the expressivity of the ansatz and the performance of the classical optimizer. For commonly used ansatzes, gradients of the objective function decay exponentially with system size, which restricts effective VQE implementations to small or carefully structured systems \cite{mcclean2018barren}. Moreover, VQE typically embeds constraints in the objective via penalty terms, resulting in challenges similar to QA~\cite{ryabinkin2018constrained}.

IQGA is an evolutionary algorithm that represents candidate solutions using quantum chromosomes and updates them by rotating gene amplitudes toward the best solution found so far. This representation enables compact encoding of large solution spaces and supports parallel exploration. However, IQGA is a metaheuristic and does not guarantee global optimality, and its performance depends on careful tuning of rotation angles to balance exploration and exploitation. In addition, constraints are enforced through penalty terms, and selecting appropriate penalty coefficients introduces challenges similar to those encountered in QA and VQE~\cite{sabaawi2023quantum}.

Overall, existing quantum approaches to the CSLP are mostly heuristic, and their reported benefits depend strongly on modeling choices and hyperparameter tuning. To address this limitation, we adopt the Grover Adaptive Search (GAS) algorithm. GAS has worst-case time complexity proportional to the square root of the search-space size, providing a quadratic speedup over classical linear-time methods such as branch and bound~\cite{boyer1998tight}. Applying GAS to the CSLP requires tailored quantum subroutines. This paper embeds the CSLP within the GAS framework by presenting two problem-specific circuits. The first identifies feasible charging station configurations, and the second evaluates their cost.}
 
\section{Problem Statement}
\label{sec:Problem Statement}
This paper addresses the challenge of identifying optimal locations for fast charging stations to facilitate long-distance trips by electric vehicles within transportation networks. Specifically, given a set of origin--destination paths, $\mathcal{Q}$, the objective is to determine a combination of locations that requires the fewest charging stations. These stations must allow EVs with a driving range of $R$ to complete each round-trip journey from origin to destination and back without running out of charge. \textcolor{black}{The set $\mathcal{Q}$ may include multiple alternative paths for the same origin-destination pair.} The vehicle is assumed to start its journey with a half-full battery, following common practice in the literature \textcolor{black}{\cite{kuby2005flow}. This ensures that the vehicle could have traveled from the nearest charging station to the origin and still return to that station if necessary.}

The transportation network is modeled as a graph $G = (\mathcal{N},E)$, where $\mathcal{N}$ is the set of nodes, representing possible locations for charging stations, and $E$ is the set of edges, representing direct connections between nodes. The length of each edge is the distance between its incident nodes. For each path \( q \in \mathcal{Q} \), \( \mathcal{N}_q \) denotes the set of nodes belonging to trip \( q \).
In addition to the network nodes, two auxiliary nodes are introduced: \( O \), which is connected to all trip origins, and \( D \), which is connected to all trip destinations. The edges connecting the auxiliary nodes to the network have zero length.

For a given node $i \in \mathcal{N}_q \cup \{O,D\}$, the accessible set $A_i^q$ comprises nodes that appear after $i$ on path $q$ and are within a distance less than the predetermined threshold value $R$ from node $i$. The first condition ensures that the vehicle does not move backward along its path. The second condition identifies the nodes that are reachable from node $i$, assuming the vehicle departs from node $i$ with a full battery. As there are no nodes after the destination, the accessible set for the destination is always empty, i.e., \( A_{D}^q = \varnothing \quad \forall q \in \mathcal{Q} \). Additionally, considering that the vehicle starts its trip with a half-full battery, the accessible nodes for node $O$ must be within a distance of $R/2$. \textcolor{black}{In this problem, there is no need to explicitly check for the return trip. As long as the electric vehicle has at least half of its battery capacity remaining when reaching the destination, the return trip is feasible using the same charging stations in the opposite direction. This is because the remaining charge is sufficient to reach the last outbound charging station and recharge there.} To ensure a successful return trip, any node whose accessible set includes node $D$ should be within a distance of $R/2$ from $D$. Let $S_i$ be a binary decision variable that equals $1$ if a charging station is installed at node $i \in \mathcal{N}$, and $0$ otherwise. This problem is formulated as follows:
\begin{subequations}
\begin{align}
\text{Minimize} \quad & \sum_{i \in \mathcal{N}} S_i \label{eq:1} \\
\text{s.t.} \quad & (1 - S_i) + \sum_{j \in A_i^q} S_j \geq 1
&& \forall q \in \mathcal{Q}, \forall i \in \mathcal{N}_q \cup \{O\} \label{eq:2} \\
& S_O = 1 && \label{eq:3} \\
& S_D = 1 && \label{eq:4} \\
& S_i \in \{0, 1\} && \forall i \in \mathcal{N} \label{eq:5}
\end{align}
\end{subequations}

The objective function \eqref{eq:1} minimizes the total number of charging stations selected. Constraint \eqref{eq:2} ensures that the vehicle will visit a charging station or the destination before running out of charge. Constraints \eqref{eq:3} and \eqref{eq:4} requires that each trip starts from the origin and ends at the destination. Finally, constraint \eqref{eq:5} Finally, constraint \eqref{eq:5} enforces the binary nature of the decision variable $S_i$. Table \ref{tab:model_parameters} in appendix \ref{app:model_parameters} summarizes the parameters used.

\section{Complexity Analysis}

This section provides a proof that the proposed CSLP in Section~\ref{sec:Problem Statement} is nondeterministic polynomial-time (NP)-hard.

The proposed CSLP can be formulated as follows: given an undirected graph $G=(\mathcal{N},E)$, and a predefined number of charging stations $C$, an induced subgraph $H$ of $G$ is a solution to the CSLP if: (1) $H$ is connected, and (2) $|V(H)| \le C$. This formulation is compatible with the quantum algorithm introduced later in this paper, which identifies an upper bound on the number of nodes with charging stations.

\begin{theorem}
\label{theorem1}
The proposed CSLP is NP-hard.
\end{theorem}

\begin{proof}
The presented proof is based on similar arguments offered by Lam et al. \cite{lam2014electric} and Conrad et al. \cite{conrad2012wildlife} with some modifications. We establish a polynomial-time reduction from the vertex cover problem to the CSLP.

Given a graph $G$, a vertex cover is a subset of vertices $\tilde{\mathcal{N}} \subseteq \mathcal{N}$ such that, for every edge $(u,v) \in E$, at least one of $u$ or $v$ belongs to $\tilde{\mathcal{N}}$. The corresponding decision version of the vertex cover problem considered here asks whether, for a given graph $G$, there exists a vertex cover $\tilde{\mathcal{N}}$ satisfying $|\tilde{\mathcal{N}}| \le C$. This problem is transformed into a connected subgraph problem by constructing a new graph $G^\prime$ based on the structure of $G$. 

The graph $G^\prime=(\mathcal{N}^\prime, E^\prime)$ is constructed as follows: for every edge in $G$, a corresponding vertex is added to $G^\prime$ to represent that edge, i.e., $\mathcal{N}^\prime=\mathcal{N} \cup E$. For every distinct pair of nodes $u, v \in \mathcal{N}$, an edge $(u,v) \in E^\prime$ is added. In addition, for every edge $(u,v) \in E$, the edges $(u,e) \in E^\prime$ and $(e,v) \in E^\prime$ are created, where $e \in \mathcal{N}^\prime$ denotes the vertex corresponding to edge $(u,v) \in G$. Accordingly, the new graph $G^\prime$ contains $|\mathcal{N}|+|E|$ vertices and $\frac{|\mathcal{N}|(|\mathcal{N}|-1)}{2}+2|E|$ edges. The cost of each node is defined as follows:

\begin{equation}
    c(v)= 
\begin{cases}
    1,& \text{if } v \in \mathcal{N}\\
    0,              & \text{otherwise}
\end{cases}
\end{equation}

\noindent To prove the theorem, we show that a solution to the vertex cover problem on $G$ with cardinality at most $C$ corresponds to a solution to the connected subgraph problem on $G^\prime$ with total cost less than or equal to $C$.

Let $H$ be an induced subgraph of $G^\prime$ that is connected and  satisfies $|V(H)| \le C$. We show that the set $\mathcal{N}^{\prime\prime} = V(H) \cap \mathcal{N}$ is a vertex cover of $G$ with $|\mathcal{N}^{\prime\prime}| \le C$. Since $\mathcal{N}^{\prime\prime} \subseteq V(H)$, it follows immediately that $|\mathcal{N}^{\prime\prime}| \le C$. To show that $\mathcal{N}^{\prime\prime}$ is a vertex cover consider any edge $(u, v) \in E$. By construction of $G^\prime$, the vertex $e \in \mathcal{N}^{\prime}$ corresponding to edge $(u, v)$ is adjacent to both $u$ and $v$. Since $H$ is connected and contains $e$, at least one of its neighboring vertices $u$ or $v$ must also belong to $V(H)$. Therefore, at least one of $u$ or $v$ is contained in $\mathcal{N}^{\prime\prime}$, implying that $\mathcal{N}^{\prime\prime}$ is a vertex cover of $G$.

We also show that if $\mathcal{N}^{\prime\prime}$ is a vertex cover of $G$ with $|\mathcal{N}^{\prime\prime}| \le C$, then one can find $H$, a subgraph of $G^{\prime}$ induced by $\mathcal{N}^{\prime\prime} \cup E$ that satisfies the connectivity and cost constraints. Clearly, $H$ has the same cost as $\mathcal{N}^{\prime\prime}$ due to the definition of cost offered above. Since $\mathcal{N}^{\prime\prime}$ is a vertex cover of $G$, for every edge $(u,v)$ in $E$, at least one of the vertices must be in $\mathcal{N}^{\prime\prime}$. This means that $H$ has an edge involving $e=(u,v)$ and a vertex in $\mathcal{N}^{\prime\prime}$. Since every pair of vertices in a subset of $H$ is connected by an edge, $H$ is also connected.

This shows a reduction from the vertex cover problem to the CSLP and shows that the solution to either problem corresponds to the solution of the other problem. Since the vertex cover problem is NP-hard, CSLP is also NP-hard.
\end{proof}

\section{Background on Grover-Based Optimization}
\label{sec:Fundamentals of Quantum Computing}
This section reviews Grover's search algorithm for unstructured search problems and Grover's Adaptive Search, an extension designed to address optimization tasks.\textcolor{black}{ A brief overview of some of the fundamental concepts that are relevant to this paper is provided in Appendix~\ref{app:quantum_prelim} for interested readers.}

\subsection{Grover's Search Algorithm }
\label{sec:grover_search}
Consider an unstructured dataset with \( N = 2^n \) elements, \textcolor{black}{\( Z_0, Z_1, \ldots, Z_{N-1} \)}, among which \( M \) elements (\( 1 \leq M \leq N \)) satisfy a Boolean condition \( f \). The goal is to find an element \( Z_v \) such that \( f(Z_v) = 1 \).

This dataset can be encoded in a quantum system. The initial state of the system, \(\ket{\psi}\), is:
\begin{equation}
\label{Eq: initial state}
    \ket{\psi} = \frac{1}{\sqrt{N}} \sum_{i=0}^{N-1} \ket{Z_i}
\end{equation}
Figure \ref{fig:Grover_1_2} illustrates that all states have equal amplitudes, thus the probability of measuring each state is the same.
\begin{figure}[h]
    \centering
    \begin{tikzpicture}[scale=0.8]
        \draw[thick, -] (0,0) -- (8,0);
        \draw[gray,dashed] (0,1) -- (8,1);

        \foreach \x in {0,2,3}
            \draw[thick] (\x,1) circle (3pt) node[above, black, font=\footnotesize] {$Z_{\x}$};
            \foreach \x in {1,4}
            \fill (\x,1) circle (3pt) node[above, black, font=\footnotesize] {$Z_{\x}$};

        \node at (6,0.9) {$\mathbf{\cdots}$};

        \fill (7,1) circle (3pt) node[above, black, font=\footnotesize] {$Z_{N-1}$};
    \end{tikzpicture}
    \caption{Initial state in Grover's algorithm. Solution states (hollow circles) and non-solution states (filled black circles) have uniform amplitudes. The dashed line indicates the mean amplitude.}
    \label{fig:Grover_1_2}
\end{figure}

Grover's algorithm transforms $\ket{\psi}$ to increase the probability of measuring a solution state. The first step is to identify the solution states using a quantum oracle $O$, which is modeled as a black box that marks solution states by flipping their phase:
\begin{equation}
    \ket{Z_i} \stackrel{O}{\longrightarrow} (-1)^{f(Z_i)}\ket{Z_i}
    \label{eq:kickback}
\end{equation}
Figure \ref{fig:Grover_2_2} illustrates the system state after the oracle is applied.
\begin{figure}[h]
    \centering
        \centering
        \begin{tikzpicture}[scale=0.8]
            \draw[thick, -] (0,0) -- (8,0);
            \draw[gray,dashed] (0,0.9) -- (8,0.9);
    
            \foreach \x in {0,2,3}
                \draw[thick] (\x,-1) circle (3pt) node[below, black, font=\footnotesize] {$Z_{\x}$};
            \foreach \x in {1,4}
                \fill (\x,1) circle (3pt) node[above, black, font=\footnotesize] {$Z_{\x}$};
    
            \node at (6,0.9) {$\mathbf{\cdots}$};
    
            \fill (7,1) circle (3pt) node[above, black, font=\footnotesize] {$Z_{N-1}$};
        \end{tikzpicture}
    \caption{State of the system after applying the oracle. The amplitudes of the solution states are inverted. The dashed line indicates the mean amplitude.}
    \label{fig:Grover_2_2}
\end{figure} 

Next, the diffuser \( 2\ket{+^n}\bra{+^n} - I \) reflects the state about the mean amplitude. As shown in Figure \ref{fig:Grover_3_2}, applying the oracle and diffuser once rotates the state toward the solution subspace. Repeating this process multiple times increases the probability of measuring a solution state. Grover's algorithm finds a solution with about \( \mathcal{O}\left(\sqrt{N/M}\right) \) oracle calls, which is a quadratic speedup over the classical \( \mathcal{O}(N/M) \) search complexity.
\begin{figure}[h]
    \centering
        \centering
        \begin{tikzpicture}[scale=0.8]
            \draw[thick, -] (0,0) -- (8,0);
            \draw[gray,dashed] (0,1) -- (8,1);
            \foreach \x in {0,2,3}
                \draw[thick] (\x,3) circle (3pt) node[above, black, font=\footnotesize] {$Z_{\x}$};
            \foreach \x in {1,4}
                \fill (\x,0.9) circle (3pt) node[above, black, font=\footnotesize] {$Z_{\x}$};
            \node at (6,0.9) {$\mathbf{\cdots}$};
            \fill (7,0.9) circle (3pt) node[above, black, font=\footnotesize] {$Z_{N-1}$};
        \end{tikzpicture}
    \caption{State of the system after applying the diffuser. All amplitudes are reflected about the mean, indicated by the dashed line.}
    \label{fig:Grover_3_2}
\end{figure}

Grover's algorithm assumes the number of solution states \( M \) is known, but this is often not true in practice. To address this, Boyer et al.\ \cite{boyer1998tight} proposed an algorithm known as quantum exponential search that works without knowing \( M \).
\FloatBarrier
\subsection{Grover's Adaptive Search}
\label{sec:grover_adaptive_search}
Grover's Adaptive Search \cite{durr1996quantum} uses quantum exponential search to solve optimization problems. Given a function \( f \) that identifies valid solutions and a cost function \( T \), the goal is to find a \( Z_\nu \) such that \( f(Z_\nu) = 1 \) and \( T(Z_\nu) \) is minimized. \textcolor{black}{The algorithm starts by choosing a threshold \(\tau\) uniformly at random from the range \(0 \leq \tau \leq N - 1\). In each iteration, it searches for solutions with costs lower than the current threshold. If a better solution is found, the threshold is updated.} The algorithm returns the true minimum with probability \(1/2\), which can be boosted by repeating the algorithm a constant number of times. Overall, the process requires \( \mathcal{O}(\sqrt{N}) \) oracle queries.
\begin{algorithm}[h]
\caption{Grover's Adaptive Search \cite{durr1996quantum}}
\begin{algorithmic}
    \State Choose a threshold index $0 \leq \tau \leq N - 1$ uniformly at random
    \While{total running time $\leq 22.5 \sqrt{N} + 1.4 \log^2 N$}
        \State Initialize the states
        \State Run exponential search with an oracle that marks $Z_j$ where $T(Z_j) < T(Z_\tau)$
        \State Store the output in $Z_{\tau'}$
        \If {$T(Z_{\tau'}) < T(Z_\tau)$}
            \State Update the threshold index $\tau \gets \tau'$
        \EndIf
    \EndWhile
    \State \Return $Z_\tau$
\end{algorithmic}
\label{alg:GAS}
\end{algorithm}
\FloatBarrier
\section{Quantum Circuit Design}
\label{sec:Quantum Circuit Design}
In Grover's search algorithm (\ref{sec:grover_search}) and Grover's adaptive search (GAS) (\ref{sec:Quantum Circuit Design}), the functions \( f \) and \( T \) are treated as general functions. To apply GAS effectively, these functions must be tailored to the problem at hand. This section presents the quantum circuit design for identifying valid combinations, that is, combinations that can support the given trips and correspond to the function $f$, and for counting the number of charging stations in each combination, which corresponds to the function $T$. These circuits are then used within GAS to identify the optimal combination.

\subsection{Identifying Valid Combinations}
\label{sec:path}
This section provides a detailed explanation of the quantum circuit design used to identify and mark valid combinations. The process is first described for a single corridor and then extended to networks with multiple origins and destinations.
 
Figure \ref{fig:Onepath} illustrates a route from node $1$ to node $4$, where nodes $1$, $2$, $3$, and $4$ are potential locations for charging stations. The electric vehicle considered in this scenario has a driving range of $100$ miles. The objective is to design a quantum circuit that identifies the valid charging station placements along this path.
\begin{figure}[h]
    \centering
    \begin{tikzpicture}[node distance={30mm},very thick, main/.style = {draw, circle, font=\scriptsize}]
    \node[main] (1) at (0,0) {$1$};
    \node[main] (2) at (2,0) {$2$};
    \node[main] (3) at (5.5,0) {$3$}; 
    \node[main] (4) at (7,0) {$4$}; 
    \begin{scope}[>={stealth[black]},
        every node/.style={fill=white,circle, font=\scriptsize},
        every edge/.style={draw=black,very thick,inner sep=1pt}]
        \path [-] (1) edge node[midway,pos = 0.4, above, pos = 0.4] {$40$ miles} (2);
        \path [-] (2) edge node[midway, above, pos = 0.4] {$70$ miles} (3);
        \path [-] (3) edge node[midway, above, pos = 0.4] {$30$ miles} (4);
    \end{scope}
    \end{tikzpicture}
    \caption{Route from node $1$ to $4$ showing candidate charging station locations and distances.}
    \label{fig:Onepath}
\end{figure}

The presence of a charging station at node \( i \) is encoded by a qubit \( \ket{S_i} \), defined as:
\begin{equation}
    \ket{S_i} = 
    \begin{cases}
        \ket{1}, & \text{if a charging station is installed at node } i,\\
        \ket{0}, & \text{otherwise}
    \end{cases}
\end{equation}

In addition to these qubits, two auxiliary qubits, \( S_O \) and \( S_D \), represent the origin and destination nodes, respectively. Figure \ref{fig:Onepath+nodes} shows the expanded route, including the origin and destination, which are marked with dashed circles. A value of \( \ket{1} \) indicates that the vehicle has reached the corresponding node. Since \( O \) and \( D \) are auxiliary nodes for nodes $1$ and $4$, respectively, their distance to those nodes is defined as zero.
\begin{figure}[h]
    \centering
    \resizebox{\columnwidth}{!}{%
    \begin{tikzpicture}[node distance={30mm},very thick, main/.style = {draw, circle, font=\normalsize}, dashedmain/.style = {draw, circle, dashed, font=\normalsize}]
        \node[main] (1) at (0,0) {$1$};
        \node[main] (2) at (2,0) {$2$};
        \node[main] (3) at (5.5,0) {$3$}; 
        \node[main] (4) at (7,0) {$4$};
        \node[dashedmain] (5) at (-2,0) {$O$};
        \node[dashedmain] (6) at (9,0) {$D$};
        \begin{scope}[>={stealth[black]},
            every node/.style={fill=white,circle, font=\normalsize},
            every edge/.style={draw=black,very thick,inner sep=1pt}]
            \path [-] (5) edge node[midway,pos=0.4, above] {$0$ miles} (1);
            \path [-] (1) edge node[midway,pos=0.4, above] {$40$ miles} (2);
            \path [-] (2) edge node[midway, above] {$70$ miles} (3);
            \path [-] (3) edge node[midway, above] {$30$ miles} (4);
            \path [-] (4) edge node[midway,pos=0.4, above] {$0$ miles} (6);
        \end{scope}
    \end{tikzpicture}
    }
    \caption{Expanded route with the origin ($O$) and destination ($D$) nodes shown as dashed circles.}
    \label{fig:Onepath+nodes}
\end{figure}

The accessible sets for each node in the example corridor are as follows (see Section~\ref{sec:Problem Statement} for the definition of accessible sets): \( A_O^{corr} = \{1, 2\} \), \( A_1^{corr} = \{2\} \), \( A_2^{corr} = \{3, 4\} \), \( A_3^{corr} = \{4, D\} \), \( A_4^{corr} = \{D\} \), and \( A_D^{corr} = \emptyset \).

This example consists of $2^4$ possible combinations for placing charging stations, some of which can be valid. Each combination is represented as a quantum state of the form
\[\ket{S_{D}\:S_{4}\:S_{3}\:S_{2}\:S_{1}\:S_{O}}\]
A state $\ket{S_{D}\:S_{4}\:S_{3}\:S_{2}\:S_{1}\:S_{O}}$ is considered a valid solution if it starts at the origin and ends at the destination, that is, if $\ket{S_{O}} = \ket{S_{D}} = \ket{1}$, and does not contain an isolated charging station. An isolated charging station is defined as a station at which a vehicle charges but exhausts its battery before reaching the next charging station. Specifically, if $\ket{S_i} = \ket{1}$, then at least one element in the accessible set of node $i$ must also be in the state $\ket{1}$.

In the following, the design details of a quantum circuit for marking valid combinations are described. The circuit consists of four main steps: (1) initializing qubits, (2) identifying combinations containing isolated nodes, (3) distinguishing valid states, and (4) reversing the operations and resetting ancilla qubits to $\ket{0}$ so that they can be reused.

\subsubsection{Initialization}
IBM's Qiskit software \cite{javadi2024quantum} is used to build and simulate the quantum circuits due to its comprehensive and user-friendly framework. In Qiskit, the default state of a qubit is $\ket{0}$. To ensure that all possible combinations of charging station locations are prepared with equal probability, Hadamard gates are applied to the qubits representing candidate charging station locations. In addition, to enforce that the vehicle starts its trip at the origin and reaches the destination, the qubits $\ket{S_{O}}$ and $\ket{S_{D}}$ are set to $\ket{1}$ using the Pauli-X gates. The five ancillary qubits, denoted by \( \ket{\text{anc}} \), and the validity qubit $\ket{v_{corr}}$ remain in their initial state. The resulting quantum state after applying these gates to the initial state \( \ket{0}^{\otimes 12} \) is given by
\begin{equation}
X \otimes H^{\otimes 4} \otimes X \otimes \mathbb{I}^{\otimes 6} \ket{0}^{\otimes 12}
= \frac{1}{\sqrt{2^4}} \sum_{x \in \{0,1\}^4} \ket{1}\ket{x}\ket{1}\ket{0}^{\otimes 6}.
\end{equation}
Figure~\ref{fig:initial} illustrates the initialization stage of the oracle.
\begin{figure}[h]
    {\tiny
    \begin{quantikz}[row sep=0.4cm, column sep=0.2cm] 
    \lstick{ $S_O$: $\ket{0}$} & \gate{X} & \qw &\\
    \lstick{$S_1$: $\ket{0}$} & \gate{H} &\qw &\\
    \lstick{$S_2$: $\ket{0}$} & \gate{H} &\qw &\\
    \lstick{$S_3$: $\ket{0}$} & \gate{H} &\qw &\\
    \lstick{$S_4$: $\ket{0}$} & \gate{H} &\qw &\\
    \lstick{$S_D$: $\ket{0}$} & \gate{X} &\qw &\\
    \lstick{$anc$: $\ket{0}$}&\qwbundle{5}& \qw &\\
    \lstick{$v_{corr}$: $\ket{0}$}&  \qw& \qw& \qw \\
    \end{quantikz}}
    \caption{Qubits Initialization. \(\ket{S_O}\) and \(\ket{S_D}\) are initialized to \(\ket{1}\). \(\ket{S_i}\) for \(i \in \{1,2,3,4\}\) are initialized to \(\ket{+}\). The ancillary qubits and the validity qubit \(\ket{v_{corr}}\) are intialized to \(\ket{0}\).}
    \label{fig:initial}
\end{figure}

\subsubsection{Isolated Charging Station Detection} 

After initialization, the circuit evaluates each node to detect isolated charging stations and records this information using ancillary qubits. For a corridor with a total of $|\mathcal{N}|$ candidate places for charging stations, $|\mathcal{N}|+1$ ancillary qubits are required: one for each node and an additional one for the origin node $O$. The destination node is excluded from checking, as its accessible set is always empty.

Figure \ref{fig:isolated} illustrates part of the oracle circuit that checks whether each node is isolated or not. For node $O$, nodes $1$ and $2$ lie within its accessible range. A multi-qubit controlled-NOT gate flips the first ancillary qubit $\text{anc}_1$ when node $O$ is in state $\ket{1}$ and all of its accessible nodes are in state $\ket{0}$. Thus, an ancillary qubit in state $\ket{1}$ indicates that the corresponding node is isolated. After this operation, these nodes are restored to their original states by applying the \(X\) gate once more.
\begin{figure}[h]
    \centering
    \resizebox{\columnwidth}{!}{%
    \begin{quantikz}[row sep=0.2cm, column sep=0.2cm]
    \lstick{$S_O$:}& \qw & \ctrl{1} & \qw & \qw & \qw& \qw &\qw& \qw & \qw &\qw &\qw &\qw &\qw&\qw&\qw&\\
    \lstick{$S_1$:}& \gate{X} & \ctrl{1}& \gate{X} & \qw & \ctrl{1}& \qw & \qw& \qw & \qw &\qw &\qw &\qw &\qw&\qw&\qw&\\
    \lstick{$S_2$:}&  \gate{X} & \ctrl{4}& \gate{X} & \gate{X} &\ctrl{5}& \gate{X}  & \qw &\ctrl{1}& \qw &\qw &\qw &\qw &\qw&\qw&\qw&\\
    \lstick{$S_3$:}&  \qw & \qw & \qw & \qw& \qw& \qw&\gate{X} & \ctrl{1}&\gate{X} & \qw &\ctrl{1}&\qw &\qw&\qw&\qw&\\
    \lstick{$S_4$:}&  \qw & \qw & \qw & \qw& \qw& \qw&\gate{X}&\ctrl{4}&\gate{X} &\gate{X} & \ctrl{1}&\gate{X}&\qw & \ctrl{1}&\qw&\\
    \lstick{$S_D$:}&  \qw & \qw & \qw & \qw& \qw& \qw & \qw & \qw & \qw &\gate{X}&\ctrl{4}&\gate{X}&\gate{X}& \ctrl{5}&\gate{X}&\\
    \lstick{$anc_1$:}& \qw &\targ{}& \qw & \qw& \qw& \qw & \qw& \qw & \qw & \qw &\qw &\qw &\qw&\qw&\qw&\\
    \lstick{$anc_2$:}& \qw &\qw& \qw & \qw& \targ{}& \qw & \qw& \qw & \qw & \qw &\qw &\qw &\qw&\qw&\qw&\\
    \lstick{$anc_3$:}& \qw &\qw& \qw & \qw& \qw& \qw & \qw&\targ{}& \qw & \qw & \qw &\qw &\qw&\qw&\qw&\\
    \lstick{$anc_4$:}& \qw &\qw& \qw & \qw& \qw& \qw & \qw&\qw& \qw & \qw &\targ{} & \qw  &\qw&\qw&\qw&\\
    \lstick{$anc_5$:}& \qw &\qw& \qw & \qw& \qw& \qw & \qw &\qw & \qw & \qw &\qw & \qw &\qw &\targ{}&\qw&
    \end{quantikz}}
    \caption{Isolated Charging Station Detection. An ancillary qubit is set to $\ket{1}$ when the corresponding node is isolated.}
    \label{fig:isolated}
\end{figure}

\subsubsection{Labeling Valid Combinations}
A combination is considered valid if it contains no isolated charging stations, that is, if all ancillary qubits are $\ket{0}$. Figure~\ref{fig:labeling} illustrates the labeling process. The ancillary qubits are inspected to verify that none of them has been flipped to $\ket{1}$. If this condition is satisfied, the combination is marked as valid, and the validity qubit is flipped using a multi-qubit controlled-NOT gate. Subsequently, the ancillary qubits are returned to their original state by applying the $X$ gate.

\begin{figure}[h]
    {\tiny
    \begin{quantikz}[row sep=0.2cm, column sep=0.2cm]
    \lstick{$anc_1$:}& \gate{X}&\ctrl{5}& \gate{X}& \qw\\
    \lstick{$anc_2$:}& \gate{X}&\ctrl{4}&\gate{X}& \qw\\
    \lstick{$anc_3$:}& \gate{X}&\ctrl{3}& \gate{X}& \qw\\
    \lstick{$anc_4$:}& \gate{X}&\ctrl{2}& \gate{X}& \qw\\
    \lstick{$anc_5$:}& \gate{X}&\ctrl{1}& \gate{X}& \qw\\
    \lstick{$v_{corr}$: }& \qw & \targ{}& \qw & \qw\\
    \end{quantikz}}
    \caption{Labeling Valid Combinations. The validity qubit \(\ket{v_{corr}}\) is flipped when all ancillary qubits are \(\ket{0}\).}
    \label{fig:labeling}
\end{figure}

\subsubsection{Restoration of Ancillary Qubits}
The restoration of ancillary qubits returns them to the state $\ket{0}$. This allows the qubits to be reused in subsequent steps of the algorithm, thereby reducing the total number of qubits required. Restoration is achieved by reversing the operations in the isolation detection step, that is, by applying the gates in Figure \ref{fig:isolated} in reverse order. Figure \ref{fig:restoration} illustrates the circuit for this restoration.
\begin{figure}[h]
    \centering
    \resizebox{\columnwidth}{!}{%
    \begin{quantikz}[row sep=0.2cm, column sep=0.2cm]
    \lstick{$S_O$:}& \qw &\qw&\qw&\qw&\qw&\qw&\qw&\qw&\qw&\qw&\qw&\qw& \qw & \ctrl{1} & \qw & \qw\\
    \lstick{$S_1$:}& \qw & \qw & \qw & \qw & \qw & \qw & \qw & \qw & \qw & \qw & \ctrl{1} & \qw & \gate{X} & \ctrl{1} & \gate{X}& \qw  \\
    \lstick{$S_2$:}& \qw & \qw & \qw & \qw & \qw & \qw & \qw & \ctrl{1} & \qw & \gate{X} & \ctrl{5} & \gate{X} & \gate{X} & \ctrl{4} & \gate{X} & \qw\\
    \lstick{$S_3$:}& \qw & \qw & \qw & \qw & \ctrl{1} & \qw & \gate{X}& \ctrl{1} & \gate{X} & \qw & \qw & \qw & \qw & \qw & \qw & \qw\\
    \lstick{$S_4$:}& \qw & \ctrl{1} & \qw & \gate{X} & \ctrl{1} & \gate{X} & \gate{X} & \ctrl{4} & \gate{X} & \qw & \qw & \qw & \qw & \qw & \qw & \qw\\
    \lstick{$S_D$:}& \gate{X} & \ctrl{5} & \gate{X} & \gate{X} & \ctrl{4} & \gate{X} & \qw & \qw & \qw & \qw & \qw & \qw & \qw & \qw & \qw & \qw\\
    \lstick{$anc_1$:}& \qw & \qw & \qw & \qw & \qw & \qw & \qw & \qw & \qw & \qw & \qw & \qw & \qw & \targ{} & \qw & \qw\\
    \lstick{$anc_2$:}& \qw & \qw & \qw & \qw & \qw & \qw & \qw & \qw & \qw & \qw & \targ{} & \qw & \qw & \qw & \qw & \qw\\
    \lstick{$anc_3$:}& \qw & \qw & \qw & \qw & \qw & \qw & \qw & \targ{} & \qw & \qw & \qw & \qw & \qw & \qw & \qw & \qw\\
    \lstick{$anc_4$:}& \qw & \qw & \qw & \qw & \targ{} & \qw & \qw & \qw & \qw & \qw & \qw & \qw & \qw & \qw & \qw & \qw\\
    \lstick{$anc_5$:}& \qw & \targ{} & \qw & \qw & \qw & \qw & \qw & \qw & \qw & \qw & \qw & \qw & \qw & \qw & \qw & \qw\\
    \end{quantikz}}
    \caption{Restoration of ancillary qubits.}
    \label{fig:restoration}
\end{figure}

Integrating the previously described steps results in a quantum circuit capable of marking all valid combinations. Figure~\ref{fig:Measuring} illustrates the quantum circuit used to identify valid charging station placements for a single path.
\begin{figure}[h]
    \centering
    \resizebox{\columnwidth}{!}{%
    \begin{quantikz}[row sep=0.3cm, column sep=0.3cm]
        \lstick{ $S_O$: $\ket{0}$} & \qw & \gate[3]{\parbox{1.8cm}{\centering Initiate}}& \qw & \qw & \gate[4]{\parbox{1.8cm}{\centering Detect \\ Isolated \\ Node}} & \qw & \qw & \gate[4]{\parbox{1.8cm}{\centering Restore}} & \qw & \qw & \qw \\
        \lstick{$S_i$: $\ket{0}$} & \qwbundle{4} & \qw & \qw & \qw & \qw & \qw & \qw & \qw & \qw & \qw & \qw \\
        \lstick{$S_D$: $\ket{0}$} & \qw & \qw & \qw & \qw & \qw & \qw & \qw & \qw & \qw & \qw & \qw \\
        \lstick{$anc$: $\ket{0}$} & \qwbundle{5} & \qw & \qw & \qw & \qw & \gate[2]{\parbox{1.8cm}{\centering Label}} & \qw & \qw & \qw & \qw & \qw \\
        \lstick{$v_{corr}$: $\ket{0}$} & \qw & \qw & \qw & \qw & \qw & \qw & \qw & \qw & \qw & \qw & \qw 
    \end{quantikz}}
    \caption{ Quantum circuit for identifying valid charging station placements on a single path.}
    \label{fig:Measuring}
\end{figure}

\begin{figure}[t]
    \centering
    \includegraphics[width=\columnwidth]{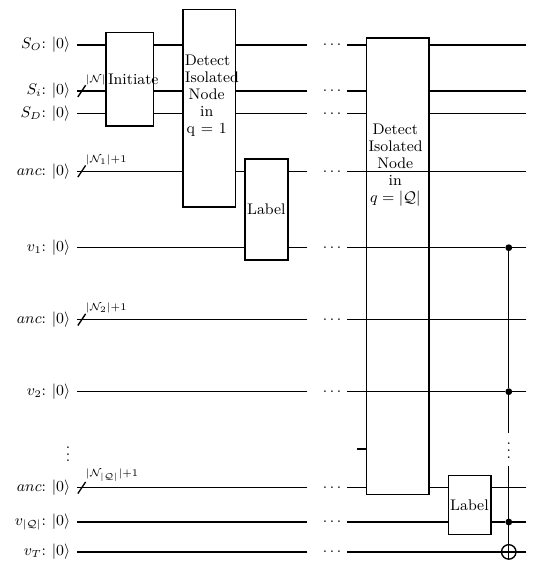}
    \caption{ Quantum circuit for identifying valid charging station placements on a network with $|\mathcal{N}|$ possible locations and $|\mathcal{Q}|$ O-D trips.}
    \label{fig:NetworkValid}
\end{figure}

\textcolor{black}{Extending the approach from a single path to a network with multiple paths is straightforward. For a network with \( |\mathcal{Q}| \) paths, all paths must be valid, that is, all validity qubits must be \( \ket{1} \). Figure~\ref{fig:NetworkValid} illustrates the corresponding quantum circuit for a network with \( |\mathcal{N}| \) nodes and \( |\mathcal{Q}| \) origin--destination pairs. In this circuit, the validity qubit associated with each path \( q \in \mathcal{Q} \) is flipped when the corresponding combination can support that path. Ultimately, the aggregation qubit \( v_T \) is flipped if all paths are traversable. The qubit restoration process is carried out by applying all gates in reverse order. The total number of qubits required is}
\begin{equation}
|\mathcal{N}| + \sum_{q \in \mathcal{Q}} |\mathcal{N}_q| + 2|\mathcal{Q}| + 3,
\end{equation}

\subsection{Station Counting}
\label{sec:station-counting}
This section extends the previously developed quantum oracle to count the number of charging stations in each combination. It then applies GAS to identify the combination with the fewest charging stations.

\textcolor{black}{Given a combination \( \ket{S_{|\mathcal{N}|}\dots S_2 S_1} \), the objective is to count the number of ones it contains. This quantity is the Hamming weight of the state and is denoted by \( \mathcal{H} = \sum_{j=1}^{|\mathcal{N}|} S_j \). We compute \( \mathcal{H} \) using an addition circuit based on Draper's addition framework~\cite{draper2000addition}.The circuit in Figure~\ref{fig:H} implements the mapping}
\begin{equation}
{\color{black}
\ket{S_{|\mathcal{N}|}\dots S_1}\ket{0}^{\otimes |\mathcal{N}|}\ \longmapsto\ \ket{S_{|\mathcal{N}|}\dots S_1}\ket{\mathcal{H}},}
\end{equation}
\textcolor{black}{In our implementation, the counting subroutine uses \( |\mathcal{N}| \) ancillary qubits as temporary workspace; however, only \( \lceil \log_2(|\mathcal{N}|+1) \rceil \) qubits store the final value of \( \mathcal{H} \), and the remaining ancillary qubits return to the state \( \ket{0} \). A proof that the proposed circuit computes the Hamming weight is provided in Appendix~\ref{app:hamming_proof}.}

\begin{figure}[h]
    \centering
    \resizebox{\columnwidth}{!}{%
    \begin{quantikz}[row sep=0.4cm, column sep=0.4cm] 
    \lstick{$S_1$} & \qw & \qw & \ctrl{4} & \ctrl{5} & \ \dots \  & \ctrl{7} & \ \dots \ & \qw  & \qw & \ \dots \ & \qw & \qw & \qw\\
    \lstick{$S_2$} & \qw & \qw & \qw & \qw & \ \dots \ & \qw  & \ \dots \ & \qw  & \qw & \ \dots \ & \qw & \qw & \qw\\
    \lstick{\vdots} & & & & & & & & & & & & & \\
    \lstick{$S_{|\mathcal{N}|}$} & \qw & \qw & \qw & \qw & \ \dots \ & \qw  & \ \dots \ & \ctrl{1} & \ctrl{2} &  \ \dots \ & \ctrl{4}  & \qw   & \qw\\
    \lstick{$\ket{0}$} & \qw & \gate{H} & \gate{R_1} & \qw & \ \dots \ & \qw & \ \dots\  & \gate{R_1} & \qw &  \ \dots \  & \qw  & \gate[4]{QFT^{\dagger}} & \qw\\
    \lstick{$\ket{0}$} & \qw & \gate{H} & \qw & \gate{R_2} & \ \dots \ & \qw  & \ \dots \ & \qw & \gate{R_2} &  \ \dots \  & \qw  & \qw & \qw\\
    \lstick{\vdots} & & & & & & & & & & & & & \\
    \lstick{$\ket{0}$} & \qw & \gate{H} & \qw & \qw & \ \dots \  & \gate{R_{|\mathcal{N}|}} & \ \dots \ & \qw  & \qw & \ \dots \ & \gate{R_{|\mathcal{N}|}} & \qw & \qw\\
    \end{quantikz}}
    \caption{Circuit for computing the Hamming weight of \(\ket{S_{|\mathcal{N}|} \dots S_1}\). The result is stored in ancillary qubits. QFT denotes the quantum Fourier transform.}
    \label{fig:H}
\end{figure}

To identify the combination with the fewest charging stations, the Hamming weights are compared against a threshold \(\tau\). This comparison is performed using Qiskit \texttt{IntegerComparator}, which checks whether each weight is less than \(\tau\):
\begin{equation}
    \ket{\mathcal{H}} \ket{0} \rightarrow \ket{\mathcal{H}} \ket{\mathcal{H} < \tau}
\end{equation}
The comparator requires \(\lceil \log(|\mathcal{N}|+1) \rceil\) ancillary qubits, equal to the number of input qubits, while storing the result in a single qubit.

\textcolor{black}{With the methodology established for identifying valid combinations and counting their charging stations, combinations with counts below specified threshold $\tau$ can be marked. The threshold $\tau$ is initialized uniformly at random from $\{1,\dots,|\mathcal{N}|+1\}$. Upon observing a valid combination with a strictly smaller charging-station count, $\tau$ is updated to that value. Building on these components, GAS can be employed to find a valid combination whose charging station count is less than $\tau$.}

Figure~\ref{fig:NewGrover} illustrates the quantum circuit for a single Grover iteration. The circuit consists of a total of 
\begin{equation}
\label{eq:num_qubit}
    |\mathcal{N}| + \sum_{q \in \mathcal{Q}} |\mathcal{N}_q| + 2|\mathcal{Q}|+ \max \{2\lceil\log(|\mathcal{N}|+1)\rceil+1, |\mathcal{N}|\}+4
\end{equation}
qubits. Of these, $|\mathcal{N}|+2$ qubits are used to represent the origin, destination, and $|\mathcal{N}|$ corridor nodes. One qubit is initialized in $\ket{-}$ and is used for phase flipping. The remaining qubits serve as ancillary qubits and are restored at the end of each iteration.

The subroutine Valid is the oracle described in Section~\ref{sec:path} and shown in Figure~\ref{fig:NetworkValid}. It identifies and marks valid combinations. After marking, the $\mathcal{H}$ subroutine counts the number of charging stations within each combination and encodes the result onto the ancillary qubits. The comparator then compares these counts against a predefined threshold, $\tau$, and marks combinations with fewer charging stations than $\tau$. A controlled-controlled-not (CCNOT) gate flips the phase of combinations that are both valid and contain fewer than $\tau$ charging stations.

The next three subroutines ($Valid^{\dagger}$, $\mathcal{H}^{\dagger}$, and $compare^{\dagger}$) undo the previous operations and restore the qubits for subsequent Grover iterations. Finally, the diffuser operator amplifies the probability of measuring the solution states by rotating them around the mean.

\begin{figure}[h]
    \centering
     {\resizebox{\columnwidth}{!}{%
     \begin{quantikz}[row sep=0.3cm, column sep=0.2cm,transparent]
         \lstick{$\ket{0}$} & \qw & \qw & \gate[5]{Valid} & \qw & \qw & \qw & \qw & \qw & \gate[5]{Valid^{\dagger}} & \gate[3]{2\ket{+^n}\bra{+^n}-I} & \qw\\
         \lstick{$\ket{0}$} & \qwbundle[alternate]{ } & \qw& \qw & \gate[6, label style = {yshift=-1.3 cm}]{\mathcal{H}} & \qw & \qw & \qw & \gate[6,label style = {yshift=-1.3cm}]{\mathcal{H^{\dagger}}} & \qw & \qw & \qw\\
         \lstick{$\ket{0}$} & \qw & \qw& \qw & \linethrough & \qw & \qw & \qw & \linethrough & \qw & \qw & \qw\\
         \lstick{$\ket{0}$} & \qwbundle[alternate]{ } & \qw& \qw & \linethrough & \qw & \qw & \qw & \linethrough & \qw & \qw & \qw\\
         \lstick{$\ket{0}$} & \qw & \qw& \qw & \linethrough & \qw & \ctrl{2} & \qw & \linethrough & \qw & \qw & \qw\\
         \lstick{$\ket{0}$} &  \qwbundle[alternate]{ } & \qw & \qw & \qw & \gate[2]{Compare} & \qw & \gate[2]{Compare^{\dagger}} & \qw & \qw & \qw & \qw\\
         \lstick{$\ket{0}$} & \qw & \qw & \qw & \qw & \qw & \ctrl{1} & \qw & \qw  & \qw & \qw & \qw\\
         \lstick{$\ket{-}$} & \qw & \qw & \qw & \qw & \qw & \targ{} & \qw & \qw & \qw & \qw & \qw\\
    \end{quantikz}}
    \caption{Quantum circuit for one iteration of Grover's Subroutine. The circuit marks valid combinations, counts charging stations, filters combinations below the threshold, resets ancillary qubits, and applies Grover's diffuser to amplify their amplitudes.}
    \label{fig:NewGrover}}
\end{figure}

Algorithm~\ref{alg:CSLPAlg} outlines the procedure for finding the optimal solutionby integrating Algorithm~\ref{alg:GAS} and the quantum circuit shown in Figure~\ref{fig:NewGrover}. The circuit is invoked as a subroutine within an iterative update of \( \tau \) to identify the configuration with the minimum number of charging stations that supports a feasible round trip. 
\begin{algorithm}[H]
\caption{GAS-Based Algorithm for CSLP}\label{alg:CSLPAlg}
\small
\begin{algorithmic}
\State \textbf{Parameters}: $\lambda = 8/7$, $T_{\max} = 22.5 \sqrt{2^{|\mathcal{N}|}} + 1.4 |\mathcal{N}|$
\State \textbf{Initialize}:\textcolor{black}{ $(\textit{query\_count}, m, \textit{current\_best}) \gets (0, 1, \text{None})$}
\State Choose $\tau$ uniformly at random from $\{1,\dots,|\mathcal{N}|+1\}$
\While{$\textit{query\_count} \leq T_{\max} $}
    \State Construct a circuit with the number of qubits given in Eq.~\eqref{eq:num_qubit}
    \State Initialize the last qubit to $\ket{-}$.
    \State Choose $j$ uniformly at random from $\{0,\dots,m-1\}$
    \State $\textit{query\_count} \mathrel{+}= |\mathcal{N}| + j$
    \State Apply $j$ iterations of the Grover subroutine (Fig.~\ref{fig:NewGrover})
    \State $i \gets$ outcome of measuring the $|\mathcal{N}|$ network-node qubits
    \If{$Valid(i)=1 \land \mathcal{H}(i)<\tau$}
        \State $\tau \gets \mathcal{H}(i)$
        \State $m \gets 1$
        \State $\textit{current\_best} \gets i $
    \Else
        \State $m \gets \min(\lambda m, \sqrt{2^{|\mathcal{N}|}})$
    \EndIf 
\EndWhile
\State \textbf{return} $\textit{current\_best}$
\end{algorithmic}
\end{algorithm}

\section{Analysis and Results}
\begin{figure}[t]
  \centering
    \includegraphics[width=0.45\textwidth]{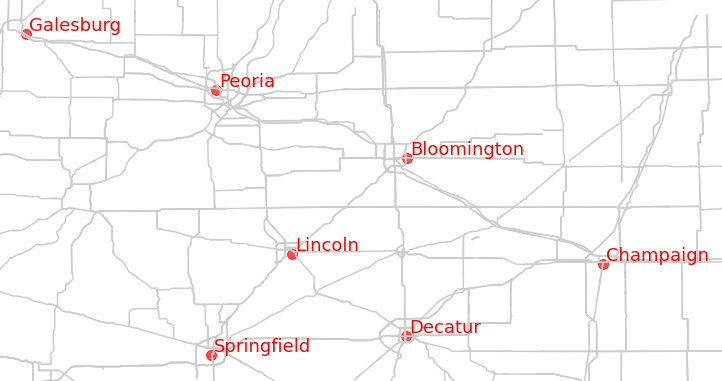}
  \caption{Map of central Illinois highlighting major roads cities.}
  \label{fig:roads}
\end{figure}

In this section, Algorithm \ref{alg:CSLPAlg} is used to determine the optimal charging station locations in the central Illinois road network, as shown in Figure \ref{fig:roads}. The goal is to facilitate travel between the major cities in central Illinois, namely Champaign, Bloomington, Decatur, Springfield, Peoria, and Galesburg. Lincoln serves as a hub in this network, though no trips originate from or are destined for this city. The map in Figure \ref{fig:roads} is converted to a network, depicted in Figure \ref{fig:Network}, where each edge represents the distance between nodes in miles. The driving range for this problem is assumed to be $R = 260$ miles. The main objective is to determine a set of charging station locations that can accommodate all trips while minimizing the total number of stations.

\begin{figure}[b]
    \centering
    \begin{tikzpicture}[scale=0.8, transform shape, node distance={30mm},very thick, main/.style = {draw, circle, font=\tiny}, dashedmain/.style = {draw, circle, dashed, font=\scriptsize}]
    \node[main] (2) at (-2,0) {$1$};
    \node[main] (3) at (-5,1.5) {$2$}; 
    \node[main] (4) at (-5,-1.5) {$3$};
    \node[main] (5) at (-6.5,0) {$4$};
    \node[main] (6) at (-7,2.5) {$5$};
    \node[main] (8) at (-9,3.5) {$6$};
    \node[main] (7) at (-7,-2.5) {$7$};

    \begin{scope}[>={stealth[black]},
        every node/.style={fill=white,circle, font=\tiny},
        every edge/.style={draw=black,very thick,inner sep=1pt}]
        \path [-] (2) edge node {$82$} (3);
        \path [-] (2) edge node {$79$} (4);
        \path [-] (3) edge node {$74$} (4);
        \path [-] (3) edge node {$61$} (6);
        \path [-] (6) edge node {$77$} (8);
        \path [-] (3) edge node {$52$} (5);
        \path [-] (5) edge node {$55$} (4);
        \path [-] (5) edge node {$55$} (7);
        \path [-] (5) edge node {$72$} (6);
        \path [-] (4) edge node {$65$} (7);
    \end{scope}
    \end{tikzpicture}
    \caption{Graph representation of central Illinois. Nodes represent cities, and edges denote intercity distances. The nodes correspond to the following cities: 1: Champaign, 2: Bloomington, 3: Decatur, 4: Lincoln, 5: Peoria, 6: Galesburg, 7: Springfield.}
    \label{fig:Network}
\end{figure}

\textcolor{black}{Although gate-based quantum computers with over $1000$ quantum bits are currently available \cite{ibmcondor}, they remain noisy, and quantum error correction is an active area of research and beyond the scope of this study. Accordingly, the IBM AER simulator is employed without modeling noise. However, this simulator imposes limitations on the size of quantum circuits. Due to this constraint, the most qubit-intensive part of Algorithm~\ref{alg:CSLPAlg}, the Valid subroutine, is implemented using the \texttt{BitFlipOracleGate} provided by the Qiskit framework.}

\textcolor{black}{Here, we clarify why, even though the \texttt{BitFlipOracleGate} is available, a custom circuit is still introduced in section~\ref{sec:path} to identify valid combinations. The \texttt{BitFlipOracleGate} relies on a generic reversible logic synthesis pipeline that is heuristic in nature and does not provide worst-case guarantees on circuit depth or on the number of ancillary qubits~\cite{shende2003synthesis}. In addition, the resulting circuit complexity depends on the syntactic structure of the Boolean expression and does not exploit problem structure. For this reason, the Valid oracle is constructed in the previous section. This construction enables a bound on the number of qubits, shown in Equation~\eqref{eq:num_qubit}, and guarantees polynomial-time circuit construction, with worst-case complexity $O(|\mathcal{Q}||\mathcal{N}|^2)$.}

\textcolor{black}{The following equation is a Boolean expression that represents the problem constraints in Equations~\eqref{eq:2}--\eqref{eq:5}. Whenever this Boolean expression evaluates to true, \texttt{BitFlipOracleGate} flips the ancillary qubit $\ket{v_T}$ from $\ket{0}$ to $\ket{1}$.}

\begin{equation}
\left( S_{O} \land S_{D}  \right) \land \bigcup_{q \in Q}\bigcup_{i \in \mathcal{N}_q }\left( \sim S_i  \lor \bigcup_{j \in A_i^q }  S_j  \right) 
\label{EQ:logicalExp}
\end{equation}

\textcolor{black}{Figure~\ref{fig:bit-flip-oracle} shows the quantum circuit obtained from \texttt{BitFlipOracleGate}, which consists of $10$ qubits. The circuit depth is $13$ when expressed using multi-controlled gates and $1859$ after decomposition into one- and two-qubit gates. Using the Valid subroutine, the resulting circuit requires $82$ qubits and has depth $3447$ under a one- and two-qubit gate decomposition.}

\begin{figure}[t]
\centering
\resizebox{\columnwidth}{!}{%
\begin{quantikz}[row sep=0.35cm, column sep=0.35cm]
\lstick{$S_O$}        & \octrl{9} & \octrl{9} & \octrl{9} & \octrl{9} & \octrl{9} & \octrl{9} & \octrl{9} & \ctrl{9}  & \ctrl{9}  & \ctrl{9}  & \ctrl{9}  & \qw       & \qw       & \qw \\
\lstick{$S_1$}  & \octrl{8} & \octrl{8} & \octrl{8} & \octrl{8} & \ctrl{8}  & \qw       & \qw       & \ctrl{8}  & \ctrl{8}  & \qw       & \qw       & \octrl{8} & \ctrl{8}  & \qw \\
\lstick{$S_2$}& \octrl{7} & \octrl{7} & \octrl{7} & \ctrl{7}  & \qw       & \ctrl{7}  & \ctrl{7}  & \ctrl{7}  & \ctrl{7}  & \ctrl{7}  & \ctrl{7}  & \ctrl{7}  & \qw       & \qw \\
\lstick{$S_3$}    & \octrl{6} & \qw       & \qw       & \octrl{6} & \octrl{6} & \octrl{6} & \ctrl{6}  & \octrl{6} & \octrl{6} & \ctrl{6}  & \ctrl{6}  & \ctrl{6}  & \ctrl{6}  & \qw \\
\lstick{$S_4$}    & \octrl{5} & \qw       & \qw       & \qw       & \qw       & \qw       & \qw       & \ctrl{5}  & \qw & \octrl{5} & \ctrl{5}  & \qw       & \qw       & \qw \\
\lstick{$S_5$}     & \octrl{4} & \octrl{4} & \ctrl{4}  & \ctrl{4}  & \ctrl{4}  & \octrl{4} & \octrl{4} & \octrl{4} & \ctrl{4}  & \octrl{4} & \octrl{4} & \ctrl{4}  & \ctrl{4}  & \qw \\
\lstick{$S_6$}& \octrl{3} & \qw       & \qw       & \ctrl{3}  & \ctrl{3}  & \ctrl{3}  & \qw       & \ctrl{3}  & \ctrl{3}  & \ctrl{3}  & \qw       & \qw       & \qw       & \qw \\
\lstick{$S_7$}  & \octrl{2} & \ctrl{2}  & \qw       & \qw       & \qw       & \ctrl{2}  & \ctrl{2}  & \ctrl{2}  & \qw       & \ctrl{2}  & \ctrl{2}  & \qw       & \qw       & \qw \\
\lstick{$S_D$}        & \qw       & \ctrl{1}  & \ctrl{1}  & \ctrl{1}  & \ctrl{1}  & \ctrl{1}  & \ctrl{1}  & \ctrl{1}  & \ctrl{1}  & \ctrl{1}  & \ctrl{1}  & \ctrl{1}  & \ctrl{1}  & \qw \\
\lstick{$v_T$}        & \targ{}   & \targ{}   & \targ{}   & \targ{}   & \targ{}   & \targ{}   & \targ{}   & \targ{}   & \targ{}   & \targ{}   & \targ{}   & \targ{}   & \targ{}   & \qw \\
\end{quantikz}}
\caption{Quantum circuit generated by \texttt{BitFlipOracleGate} for $R=260$.}
\label{fig:bit-flip-oracle}
\end{figure}

After constructing the Valid subroutine, the remaining steps follow the same process as described in section~\ref{sec:station-counting}. \textcolor{black}{Over 314 independent runs of Algorithm~2, the algorithm reached the global minimum in 30.2\% of the experiments.} Figure~\ref{fig:ev_subfigs} illustrates the two optimal charging-station configurations corresponding to the minimum station count.
\begin{figure*}[t]
\centering

\begin{minipage}[t]{0.48\textwidth}
\centering
\subfloat[]{
\begin{tikzpicture}[scale=0.7, very thick,
  main/.style = {draw, circle, minimum size=0.6cm, font=\tiny},
  ev/.style   = {draw, rectangle, rounded corners, minimum width=0.5cm,
                 minimum height=0.8cm, font=\tiny, fill=gray!20}]
  \node[main] (2) at (-2,0) {$1$};
  \node[ev]   (3) at (-5,1.5) {$2$};
  \node[ev]   (4) at (-5,-1.5) {$3$};
  \node[main] (5) at (-6.5,0) {$4$};
  \node[ev]   (6) at (-7,2.5) {$5$};
  \node[main] (8) at (-9,3.5) {$6$};
  \node[main] (7) at (-7,-2.5) {$7$};

  \begin{scope}[>={stealth[black]},
    every node/.style={fill=white,circle, font=\tiny},
    every edge/.style={draw=black,very thick,inner sep=1pt}]
    \path [-] (2) edge node {$82$} (3);
    \path [-] (2) edge node {$79$} (4);
    \path [-] (3) edge node {$74$} (4);
    \path [-] (3) edge node {$61$} (6);
    \path [-] (6) edge node {$77$} (8);
    \path [-] (3) edge node {$52$} (5);
    \path [-] (5) edge node {$55$} (4);
    \path [-] (5) edge node {$55$} (7);
    \path [-] (5) edge node {$72$} (6);
    \path [-] (4) edge node {$65$} (7);
  \end{scope}
\end{tikzpicture}
\label{fig:ev_2_3_5}}
\end{minipage}
\hfill
\begin{minipage}[t]{0.48\textwidth}
\centering
\subfloat[]{
\begin{tikzpicture}[scale=0.7, very thick,
  main/.style = {draw, circle, minimum size=0.6cm, font=\tiny},
  ev/.style   = {draw, rectangle, rounded corners, minimum width=0.5cm,
                 minimum height=0.8cm, font=\tiny, fill=gray!20}]
  \node[ev]   (2) at (-2,0) {$1$};
  \node[main] (3) at (-5,1.5) {$2$};
  \node[ev]   (4) at (-5,-1.5) {$3$};
  \node[main] (5) at (-6.5,0) {$4$};
  \node[ev]   (6) at (-7,2.5) {$5$};
  \node[main] (8) at (-9,3.5) {$6$};
  \node[main] (7) at (-7,-2.5) {$7$};

  \begin{scope}[>={stealth[black]},
    every node/.style={fill=white,circle, font=\tiny},
    every edge/.style={draw=black,very thick,inner sep=1pt}]
    \path [-] (2) edge node {$82$} (3);
    \path [-] (2) edge node {$79$} (4);
    \path [-] (3) edge node {$74$} (4);
    \path [-] (3) edge node {$61$} (6);
    \path [-] (6) edge node {$77$} (8);
    \path [-] (3) edge node {$52$} (5);
    \path [-] (5) edge node {$55$} (4);
    \path [-] (5) edge node {$55$} (7);
    \path [-] (5) edge node {$72$} (6);
    \path [-] (4) edge node {$65$} (7);
  \end{scope}
\end{tikzpicture}
\label{fig:ev_1_3_5}}
\end{minipage}

\caption{Optimal charging station locations. (a) Nodes 2, 3, and 5 require charging stations. (b) Nodes 1, 3, and 5 require charging stations.}
\label{fig:ev_subfigs}
\end{figure*}

\begin{figure}[h]
    \centering
    \includegraphics[width=0.9\linewidth]{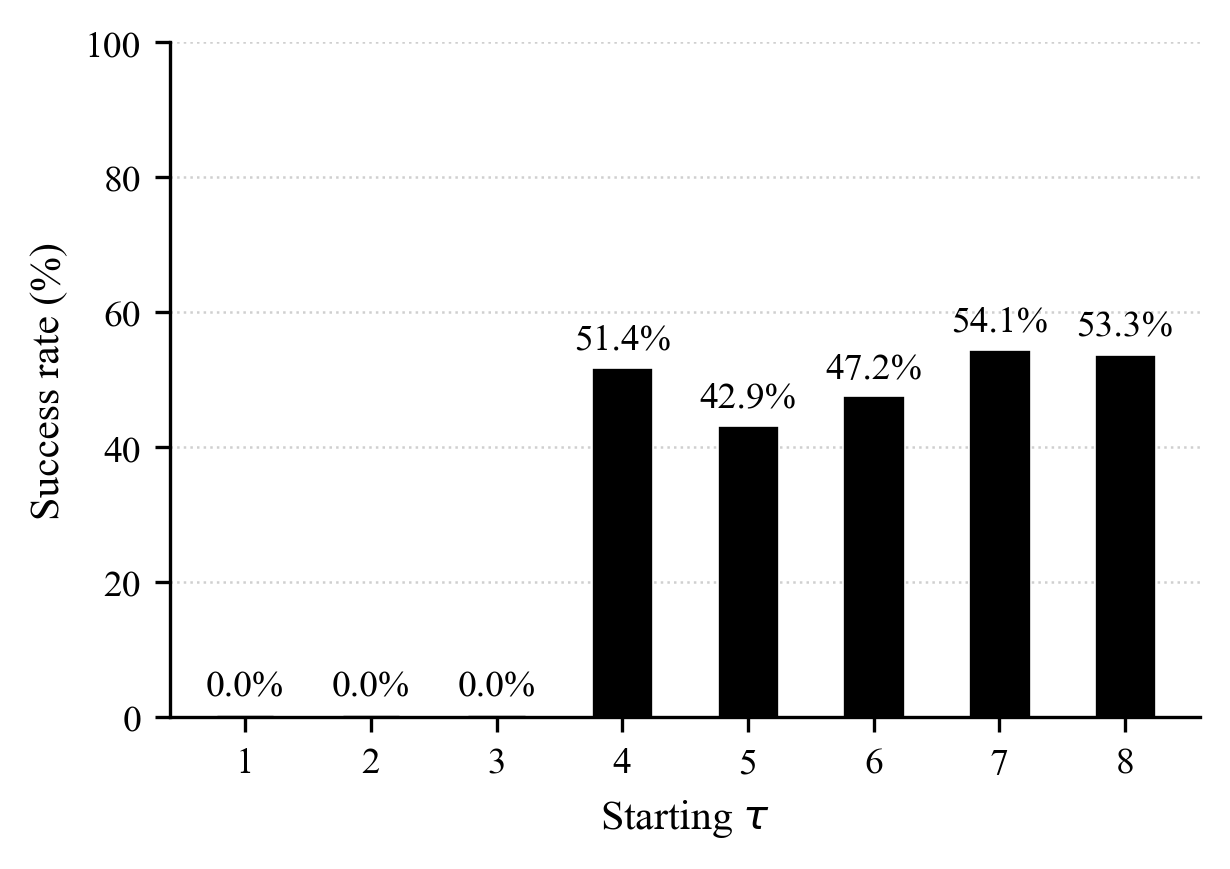}
    \caption{Success rate of Algorithm~2 versus the starting $\tau$.}
    \label{fig:histogram}
\end{figure}

\textcolor{black}{The observed success rate is below the expected $50\%$ because the problem instance is small, making the infeasible search space relatively large. The initial threshold $\tau$ is selected uniformly at random from $\{1,\ldots,8\}$, and with probability $3/8$ it is initialized below the true minimum objective value. In this case, no feasible solution satisfies $\mathcal{H}(i)<\tau$, and the algorithm terminates without success. Figure~\ref{fig:histogram} shows the success rate for different initial values of $\tau$. When $\tau$ is initialized above the minimum value, the algorithm achieves the expected success rate. To avoid the low success rate in the following experiments, the initial threshold is fixed at $\tau=8$.}

\begin{figure}[h]
    \centering
    \includegraphics[width=0.9\linewidth]{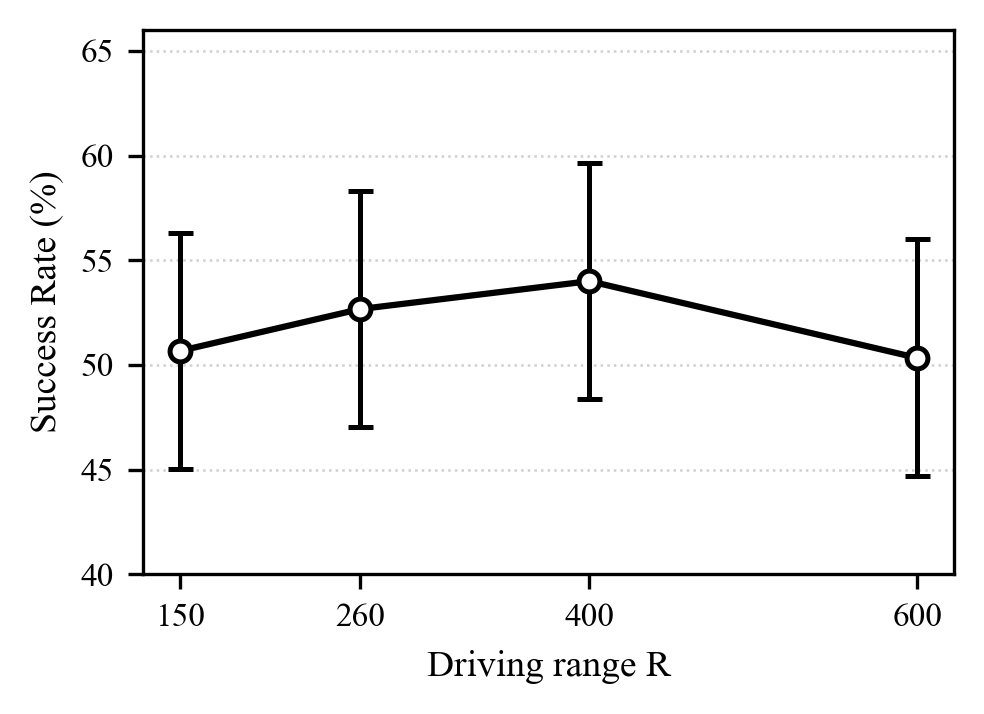}
    \caption{Sensitivity of the success rate to the driving range. Error bars denote 95\% confidence intervals.}
    \label{fig:sensitivity-R}
\end{figure}

\textcolor{black}{Figure~\ref{fig:sensitivity-R} shows the sensitivity of the success rate to the driving range. No statistically significant differences are observed across the tested values of $R$, indicating that the algorithm's performance is robust with respect to changes in the driving range.}

\begin{figure}[h]
    \centering
    \includegraphics[width=0.9\linewidth]{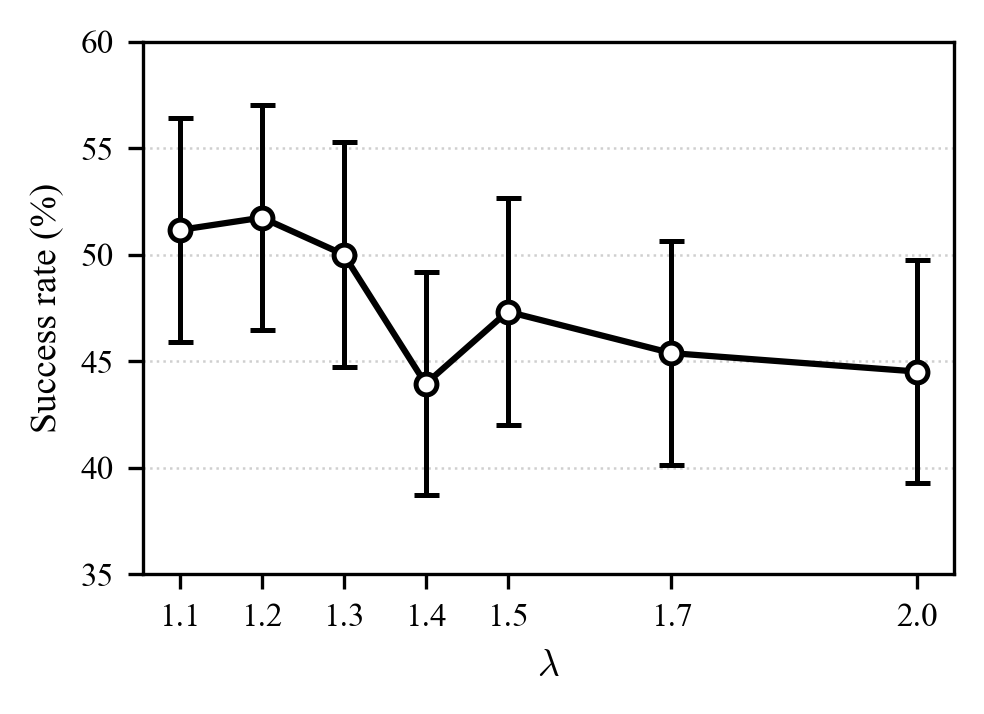}
    \caption{Success rate of Alg.~\ref{alg:CSLPAlg} with $R=260$ miles as a function of $\lambda$. Error bars denote 95\% confidence intervals.}
    \label{fig:sensitivity-lambda}
\end{figure}

\textcolor{black}{Figure~\ref{fig:sensitivity-lambda} reports the success rate for Alg.~\ref{alg:CSLPAlg} with $R = 260$ miles as a function of $\lambda$. The success rate is approximately $0.50\!-\!0.52$ for $\lambda \in \{1.1, 1.2, 1.3\}$ and decreases to approximately $0.44$ at $\lambda = 1.4$.  The success rate at $\lambda = 1.4$ is lower than that at $\lambda = 1.2$ at the 95\% confidence level. Larger values of $\lambda$ make the range from which the number of Grover rotations is chosen grow more quickly, increasing the likelihood of exceeding the optimal rotation value and reducing the success rate.}

\section{Conclusion and future path}
The placement of public EV charging stations along interstate highways is critical for enabling long-distance EV travel. Optimizing charging station locations is an NP-hard problem, which motivates the exploration of alternative computational approaches. This study introduces a quantum optimization framework based on Grover Adaptive Search to identify optimal charging station configurations for long-distance trips.

The results show that the proposed quantum circuits effectively formulate the CSLP within the Grover Adaptive Search framework, with empirical success rates consistent with theoretical expectations. The method exhibits robustness to changes in the driving range. Moreover, the selected value $\lambda = 8/7$ lies in a favorable parameter regime, whereas larger values lead to degraded performance.

{\color{black}Grover Adaptive Search is intended for fault-tolerant quantum computers and relies on deep quantum circuits. Because such circuits cannot be executed reliably on current noisy hardware, all experiments are performed on a quantum simulator with a limited number of qubits, which restricts the size of the network considered in the case study. Table~\ref{Tab:QHardware} summarizes the capabilities of current gate-based quantum hardware platforms and highlights that, despite steady progress, present devices do not yet support large, error-corrected circuits. The results of this study therefore focus on validating the correctness of the proposed formulation under current technological constraints.

The proposed quantum subroutines admit explicit bounds on both qubit count and circuit construction time, but tightening these bounds is beyond the scope of this work. Future work may focus on synthesizing more efficient quantum circuits with fewer qubits and shallower depth. GAS provides a theoretical quadratic speedup. However, the small problem size and current qubit limitations prevent evaluating this advantage empirically. Demonstrating a speedup requires varying the problem size over a wide range and observing how runtime changes with $N$. Since this is not possible here, the experiments are used to validate correctness and expected algorithmic behavior. An empirical study of the quadratic speedup on larger problem instances, where classical methods become impractical, is left for future work.} The framework can also be extended to incorporate more practical constraints, such as charging station capacity limits, congestion, or station reliability.

\begin{table*}[h]
{\color{black}
\centering
\caption{\small Universal gate-based quantum hardware platforms \cite{quantumzeitgeist2026quantum}.}
\renewcommand{\arraystretch}{1.3}
\small
\begin{tabularx}{\textwidth}{
  L{2.2cm}  
  L{1.3cm}  
  L{2.2cm}  
  L{1.4cm}  
  L{1.2cm}  
  L{1.8cm}  
  X         
}
  \toprule
  \makecell[l]{Qubit\\technology} &
  \makecell[l]{Qubit\\count} &
  \makecell[l]{Entanglement\\connectivity} &
  \makecell[l]{Two-qubit\\gate fidelity} &
  \makecell[l]{Gate\\time} &
  \makecell[l]{$T_2$ coherence\\time} &
  Notes \\
  \midrule
  Superconducting &
  100--156 &
  Nearest-neighbor &
  99.7\% &
  50 ns &
  300 \textmu{}s &
  Fast, well-controlled operations; coherence and locality constraints.
  \\
  Trapped-ion &
  50--100 &
  All-to-all &
  $\geq$99.9\% &
  1--100 \textmu{}s &
  \makecell[l]{Seconds to\\minutes} &
  High-fidelity operations; slower gates and modular scaling.
  \\
  Photonic &
  90--200 &
  Programmable &
  $\sim$99\% &
  ps--ns &
  Long &
  Room-temperature, ultrafast operations; loss and probabilistic entanglement.
  \\
  Silicon spin &
  6--12 &
  Nearest-neighbor &
  $\sim$99\% &
  1--10 ns &
  \makecell[l]{Milliseconds\\to seconds} &
  Compact, fabrication-compatible platform; control and scalability challenges.
  \\
  Neutral-atom &
  1,200+&
  Reconfigurable all-to-all &
  $\sim$99.5\% &
  1--10 \textmu{}s &
  1--13 s &
  Large, reconfigurable arrays; uniform high-fidelity control still maturing.
  \\
  \bottomrule
\end{tabularx}
\label{Tab:QHardware}}
\end{table*}

\section*{Author Contributions}
\textbf{Tina Radvand}: Conceptualization, Methodology, Formal Analysis, Writing -- Original Draft, , Writing -- Review \& Editing, Visualization, Software. \textbf{Alireza Talebpour}: Conceptualization, Funding Acquisition, Supervision, Writing -- Review \& Editing. \textbf{Homa Khosravian}: Formal Analysis.

\section*{Declaration of Generative AI}
During the preparation of this work, the authors used ChatGPT to refine sentence structure and improve clarity.

\appendices
{\color{black}
\begingroup
\section{Quantum computing Preliminaries}
\label{app:quantum_prelim}

This appendix reviews the concepts of qubit, superposition, and quantum gate. Note that while this appendix does not aim to provide a comprehensive overview of fundamental concepts in quantum computing, it covers the core concepts used in this paper. Interested readers are referred to~\cite{NielsenChuang2010} for an introduction to the principles of quantum computing.

\subsection{Qubit and Superposition}
\label{subsec:Qubit and Superposition}
A qubit is the quantum analogue of a classical bit. The computational basis states of a qubit are
$\ket{0}=\begin{bmatrix}1\\0\end{bmatrix}$ and
$\ket{1}=\begin{bmatrix}0\\1\end{bmatrix}$.
The symbol $\ket{\cdot}$ (``ket'') represents a quantum state as a column vector, and $\bra{\cdot}$ (``bra'') represents the corresponding row vector. A bit can be only $0$ or $1$, whereas a qubit can be in a linear combination of $\ket{0}$ and $\ket{1}$, i.e., $\alpha_0 \ket{0} + \alpha_1 \ket{1}$, where $\alpha_0$ and $\alpha_1$ are complex numbers.
This state is called a superposition. When a measurement is performed on this superposition, it collapses, and the result is either the state $\ket{0}$ with probability $|\alpha_0|^2$ or $\ket{1}$ with probability $|\alpha_1|^2$. Note that $|\alpha_0|^2 + |\alpha_1|^2 = 1$.Transitioning from the single-qubit system to $n$ qubits, the general quantum state can be written as
\[
\sum_{x \in \{0,1\}^n} \alpha_x \ket{x},
\]
Where $\sum_{x \in \{0,1\}^n} |\alpha_x|^2 = 1$.
Superposition allows quantum systems to simultaneously exist in multiple states, in contrast to classical systems that are limited to a single state. This property enables quantum parallelism: when a function is applied to the system, it acts on all states simultaneously.

\subsection{Quantum Gates}
Quantum gates are operators that act on qubits to transform their quantum state. The gates used in this paper are summarized in Table \ref{tab:gates_simple}.

\begin{table}[h]
\centering
{\color{black}
\caption{\small Summary of quantum gates used in this paper.}
\footnotesize
\setlength{\tabcolsep}{6pt}
\renewcommand{\arraystretch}{1.2}
\begin{tabularx}{\columnwidth}{l X}
\toprule
Gate & Action \\
\midrule
Pauli-X ($X$)
& Flips the qubit state: $X\ket{0}=\ket{1}$ and $X\ket{1}=\ket{0}$. \\[1.2mm]

Hadamard ($H$)
& Converts $\ket{0}$ and $\ket{1}$ into equal-weight combinations of $\ket{0}$ and $\ket{1}$:
    $H\ket{0}=\dfrac{\ket{0}+\ket{1}}{\sqrt{2}}$ and
    $H\ket{1}=\dfrac{\ket{0}-\ket{1}}{\sqrt{2}}$. \\[1.2mm]

CNOT (CNOT)
& Two-qubit gate that flips the target when the control is $\ket{1}$:
  $\ket{00}\!\mapsto\!\ket{00}$, $\ket{01}\!\mapsto\!\ket{01}$, $\ket{10}\!\mapsto\!\ket{11}$, $\ket{11}\!\mapsto\!\ket{10}$. The first qubit is called the control qubit. \\[1.2mm]

Phase ($R_n$)
& Adds a phase to the $\ket{1}$ component:
  $R_n\ket{0}=\ket{0}$ and $R_n\ket{1}=e^{\frac{2\pi i}{2^n}}\ket{1}$. \\
\bottomrule
\end{tabularx}
\label{tab:gates_simple}}
\end{table}

\section{Model Parameters}
\label{app:model_parameters}

Table~\ref{tab:model_parameters} summarizes the sets, parameters, and decision variables used in this paper.
\begin{table}[h]
    \caption{\small Model Parameters}
    \centering
    \small
    \begin{tabularx}{\columnwidth}{>{\hsize=0.3\hsize}X >{\hsize=1.5\hsize}X}
        \hline
        \textbf{Parameter}   & \textbf{Description} \\ \hline
        $\mathcal{Q}$        & Set of origin--destination trips\\
        $q$                  & A specific trip, where $q \in \mathcal{Q}$ \\ 
        $R$                  & Driving range of an electric vehicle   \\
        $A_i^q$              & Set of accessible nodes for node $i$ on path $q$\\
        $\mathcal{N}$        & Set of nodes of network $G$\\
        $E$                  & Set of edges of network $G$\\
        $\mathcal{N}_q$      & Set of nodes associated with trip $q$\\
        $O$                  & Auxiliary node connected to all trip origins \\
        $D$                  & Auxiliary node connected to all trip destinations \\

        \multicolumn{2}{l}{\textbf{Decision Variables}}\\
        $S_i$                & Binary station location variable; $S_i =1$ if station is located at node $i$, $S_i =0$ otherwise   \\

        \hline
    \end{tabularx}
    \label{tab:model_parameters}
\end{table}

\section{Proof of Correctness for the Hamming-Weight Circuit}
\label{app:hamming_proof}

We show that the proposed circuit (Figure~\ref{fig:H}) implements the mapping
\[
\ket{S_{|\mathcal{N}|} \dots S_1} \ket{0}^{\otimes |\mathcal{N}|}
\;\mapsto\;
\ket{S_{|\mathcal{N}|} \dots S_1} \ket{\mathcal{H}}.
\]

\begin{proof}
Let
\begin{equation}
 |\psi_0\rangle =
\ket{S_{|\mathcal{N}|} \dots S_1} \ket{0}^{\otimes |\mathcal{N}|}.
\end{equation}

After applying Hadamard gates to each ancillary qubit, the state becomes
\begin{equation}
|\psi_1\rangle
=
\ket{S_{|\mathcal{N}|} \dots S_1}
\otimes
\frac{1}{\sqrt{2^{|\mathcal{N}|}}}
\sum_{y=0}^{2^{|\mathcal{N}|}-1} \ket{y},
\end{equation}
where \( \ket{y} = \ket{y_1 y_2 \dots y_{|\mathcal{N}|}} \). When the circuit applies the controlled rotation gates
\( R_1, R_2, \dots, R_{|\mathcal{N}|} \),
the basis state \( \ket{y} \) acquires the phase
\begin{equation}
\prod_{k=1}^{|\mathcal{N}|}
\left( e^{2\pi i / 2^k} \right)^{y_k S_j}
=
e^{2\pi i S_j \sum_{k=1}^{|\mathcal{N}|} y_k / 2^k}
\end{equation}
for each input bit \( S_j \). Applying this operation for each input bit \( S_j \) yields the cumulative phase
\begin{equation}
e^{2\pi i
\left( \sum_{j=1}^{|\mathcal{N}|} S_j \right)
\left( \sum_{k=1}^{|\mathcal{N}|} y_k / 2^k \right)}
=
e^{2\pi i \, \mathcal{H}
\sum_{k=1}^{|\mathcal{N}|} y_k / 2^k}.
\end{equation}
Thus, the state of the system becomes
\begin{equation}
|\psi_2\rangle
=
\ket{S_{|\mathcal{N}|} \dots S_1}
\otimes
\frac{1}{\sqrt{2^{|\mathcal{N}|}}}
\sum_{y=0}^{2^{|\mathcal{N}|}-1}
e^{2\pi i \, \mathcal{H}
\left( \sum_{k=1}^{|\mathcal{N}|} y_k / 2^k \right)}
\ket{y}.
\end{equation}

Recall that the quantum Fourier transform satisfies
\begin{equation}
\text{QFT} \ket{x}
=
\frac{1}{\sqrt{2^{|\mathcal{N}|}}}
\sum_{y=0}^{2^{|\mathcal{N}|}-1}
e^{2\pi i \, x y / 2^{|\mathcal{N}|}} \ket{y}.
\end{equation}
Moreover, using the binary expansion of \( y \), we have
\begin{equation}
\sum_{k=1}^{|\mathcal{N}|} \frac{y_k}{2^k}
=
\frac{
y_1 \cdot 2^{|\mathcal{N}|-1}
+ y_2 \cdot 2^{|\mathcal{N}|-2}
+ \cdots
+ y_{|\mathcal{N}|} \cdot 2^0
}{2^{|\mathcal{N}|}}
=
\frac{y}{2^{|\mathcal{N}|}}.
\end{equation}

Substituting this identity into the exponent shows that
\begin{equation}
|\psi_2\rangle
=
\ket{S_{|\mathcal{N}|} \dots S_1}
\otimes
\text{QFT} \ket{\mathcal{H}}.
\end{equation}

Applying the inverse quantum Fourier transform to the ancillary register gives
\begin{equation}
|\psi_3\rangle
=
(I \otimes \text{QFT}^\dagger) |\psi_2\rangle
=
\ket{S_{|\mathcal{N}|} \dots S_1} \ket{\mathcal{H}}.
\end{equation}
This completes the proof.
\end{proof}
\endgroup}

\bibliographystyle{IEEEtran}
\bibliography{biblio}

 


\vfill

\end{document}